\documentclass[sigconf]{acmart}

\usepackage[ruled,linesnumbered]{algorithm2e}
\usepackage{algpseudocode}
\SetKw{KwBy}{by}
\SetKwComment{Comment}{/* }{ */}
\RestyleAlgo{ruled}
\usepackage{caption}
\usepackage{subcaption}

\usepackage{gensymb}

\usepackage{array,multirow,graphicx}

\AtBeginDocument{%
  \providecommand\BibTeX{{%
    \normalfont B\kern-0.5em{\scshape i\kern-0.25em b}\kern-0.8em\TeX}}}

\acmConference[]{USC FLS Lab}{flslab.org}{Private Copy}

\newtheorem{problem}{Problem Definition}

\copyrightyear{2023}
\acmYear{2023}
\setcopyright{rightsretained}
\acmConference[MMAsia '23]{ACM Multimedia Asia 2023}{December 6--8, 2023}{Tainan, Taiwan}
\acmBooktitle{ACM Multimedia Asia 2023 (MMAsia '23), December 6--8, 2023, Tainan, Taiwan}\acmDOI{10.1145/3595916.3626460}
\acmISBN{979-8-4007-0205-1/23/12}

\begin{document}

\title{
An Evaluation of Decentralized Group Formation Techniques for Flying Light Specks
}

\author{Hamed Alimohammadzadeh, Heather Culbertson, Shahram Ghandeharizadeh}
\affiliation{%
\institution{Computer Science Department, University of Southern California}
  \streetaddress{Computer Science Department}
  \city{Los Angeles}
  \state{California}
  \country{USA}
  \postcode{90089}
}
\email{{halimoha,hculbert,shahram}@usc.edu}

\acmConference[]{Multimedia Asia 2023}{Tainan, Taiwan}{Dec 6-8}


\begin{abstract}
Group formation is fundamental for 3D displays that use Flying Light Specks, FLSs, to illuminate shapes and provide haptic interactions.  An FLS is a drone with light sources that illuminates a shape.  Groups of $G$ FLSs may implement reliability techniques to tolerate FLS failures, provide kinesthetic haptic feedback in response to a user's touch, and facilitate a divide and conquer approach to challenges such as localizing FLSs to render a shape.
This paper evaluates four decentralized techniques to form groups.  An FLS implements a technique autonomously using asynchronous communication and without a global clock.  We evaluate these techniques using synthetic point clouds with known optimal solutions and real point clouds.
Obtained results show a technique named Random Subset (RS) is superior when constructing small groups (G$\leq$5) while a different technique named Closest Available Neighbor First (CANF) is superior when constructing large groups (G$\geq$10).

\end{abstract}

\maketitle

\section{Introduction}

\begin{figure}
\centering
\includegraphics[width=\columnwidth]{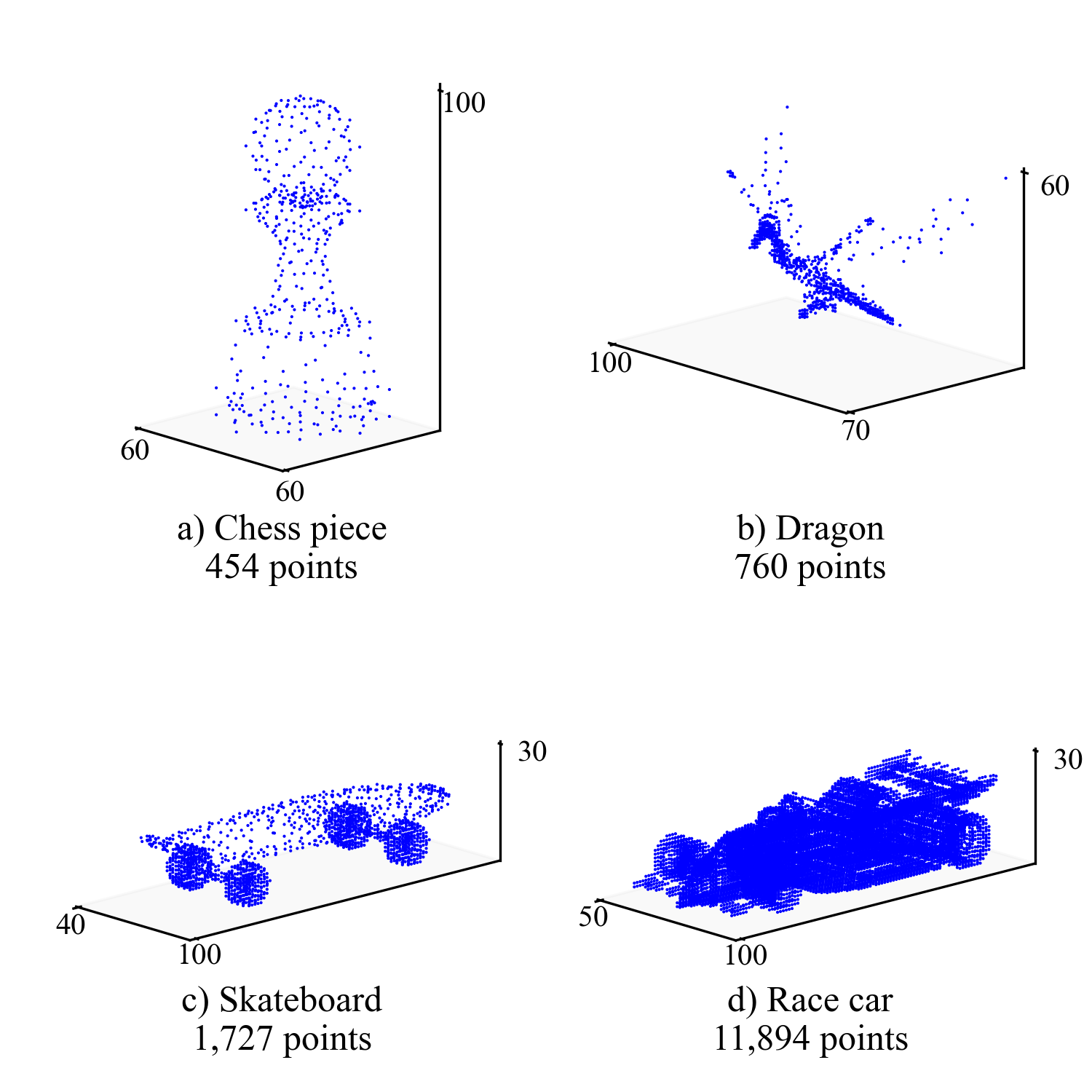}\hfill
\caption{Four different point clouds.}
\label{fig:4_shapes}
\end{figure}

A Flying Light Speck (FLS) is a miniature sized drone configured with light sources, processing, storage, and networking capabilities. 
Swarms of cooperating FLSs will illuminate complex 2D and 3D shapes in a fixed volume, an FLS display~\cite{shahram2021,shahram2022,mmsys2023}.
They will provide a user with encounter-type haptic interactions~\cite{rodrigo2021} by detecting their disposition due to the user's exerted force (e.g., poking).
In response, they provide the appropriate behavior (e.g., the shape moves away), and exert force back if necessary (e.g., emulate surface friction or hardness).  

Applications of an FLS display are diverse ranging from entertainment to health care.
With entertainment, an FLS display will illuminate characters of a multi player game such as Minecraft and Street Fighter, enabling a user to experience them from different angles.
With health care, a physician can examine a patient's MRI scans in real time by separating and analyzing the different organs (illuminations).
The physician will be able to evaluate the softness of tissue emulated by FLSs using the density readings of a scan.

An illumination is a rendering of a point cloud with an FLS assigned the coordinates of a point in the point cloud.
The illumination may require a large number (thousands if not millions) of FLSs.
Group formation is important to develop effective divide and conquer techniques for FLS displays.
Its objective is to minimize the distance between the FLSs that constitute each group.
A solution to this problem may either be centralized or decentralized.  
This paper implements and evaluates several known decentralized group formation techniques~\cite{kclique2014}.
We focus on decentralized techniques because they enable FLSs to synergize in an online manner.
Below, we describe three use cases of such an algorithm.
With each, the time to identify groups should be minimized. 

\noindent{\bf Three use cases of a decentralized group formation technique:}
With a haptic interaction such as poking, groups of FLSs may exert force at different parts of a fingertip.  To illustrate, consider a physician manipulating an organ (illumination).
The location of physician's fingertip will be detected by one or more FLSs.
The required stiffness dictates a value for $F$ and $G$.
Ideally, $F$ FLSs in the form of $nG$ groups (each with $G$ FLSs) will exert force back, enabling the physician to experience the stiffness of the tissue.
A decentralized technique that constructs these groups using the closest number of FLSs will enable formation of $nG$ groups quickly.

Second, with FLS failure handling, one may construct $nG$ reliability groups.  Each consists of $G+C$ FLSs.
$G$ FLSs illuminate a point cloud.
$C$ FLSs are dark standbys~\cite{shahram2022}.
This is similar to RAID~\cite{gibson} for mass storage devices.
Every time an FLS fails, an available standby substitutes for its lighting responsibility at its assigned coordinate.  The $G$ FLSs must be in close proximity to one another, enabling a standby to quickly substitute for a failed FLS.
This highlights the importance of minimizing distance between the FLSs in a group. 

Finally, relative localization will enable a group of FLSs to position themselves to illuminate a 2D or 3D multimedia shape.
The relative distance and angle between the FLSs will match those of the points in a point cloud.
There are a many techniques using different kinds of sensors such as UWB, IMU, optical, and their hybrids.
To illuminate a large point cloud, one may partition it into groups, each with $G$ FLSs.
The $G$ FLSs in each group may localize independent of the FLSs in the other groups. 
Subsequently, the different groups may localize relative to one another.
By reducing the distance between $G$ FLSs in a group, the likelihood of FLSs in different groups blocking one another is minimized.

We evaluate the group formation techniques along three dimensions:
1) Speed, how fast a technique computes groups, 
2) Quality of formed groups, the number of groups and the average distance between the FLSs in a group, and
3) Amount of work performed.
We quantify the first using the response time of a technique.
It is defined from the time a swarm of FLSs starts to execute a technique until they converge on a solution.
With the quality of formed groups, we focus on the number of constructed groups consisting of $G$ FLSs.
We do not consider smaller group sizes as usable, deferring an investigation of this topic to future work, see Section~\ref{sec:conc}.
The work performed by a group is quantified using the network bandwidth consumed by an FLS and its CPU processing.  
The latter is quantified by computing the average number of times an FLS executes a technique to converge on a solution.

The techniques require an FLS to consume network bandwidth in a bursty manner.
Moreover, while they may require a similar transmission rate from an FLS, the rate at which they require an FLS to receive data may be different.
Hence, we report the average transmission rate (Xmt BW) and receive rate (Rec BW) separately.

An ideal technique must provide a response time that meets the requirements of its target application, maximize the number of groups consisting of a pre-specified $G$ value, minimize the distance between the FLSs that constitute a group, and minimize the amount of network bandwidth and processing demanded from the individual FLSs. 
The {\bf contributions} of this paper include:
\begin{itemize}
\item A scalable technique to construct synthetic point clouds with known optimal grouping of FLSs and $G$ values.
It is used as a yardstick for the metrics of interest.  (Section~\ref{sec:optimal})


\item An implementation and evaluation of several known decentralized group formation heuristics:
Random Subset (RS)~\cite{Chmielowiecvns2010,kclique2014}, and Variable
Neighborhood Search (VNS)~\cite{vns2009}.
(Section~\ref{sec:decentralized}.)

\item Two new decentralized heuristic, Simple Random, SimpleR, and Closest Available Neighbor First, CANF.
(Sections~\ref{sec:simpler} and~\ref{sec:canf}.)

\item A discovery protocol to enable an FLS to identify its neighboring FLSs for group formation.
(Section~\ref{sec:impl}.)



\item An evaluation of the alternative decentralized techniques.  Obtained results show the execution time of these techniques is in the order of seconds and at times hours with dense point clouds requiring hundreds of FLSs.  This may be too slow for applications that require interactive response times faster than 100 milliseconds, e.g., haptics~\cite{rank2010}.
(Section~\ref{sec:cmp}.)

\item We open source our software implementations and its data set at \url{https://github.com/flyinglightspeck/Group-Formation}.



\end{itemize}
We survey related work in Section~\ref{sec:related} and offer brief conclusions and future research directions in Section~\ref{sec:conc}.

\section{Problem Statement and Optimality}\label{sec:optimal}

\begin{problem}\label{prob:grouping}
Given $F$ FLSs and a required group size $G$, maximize the number of groups consisting of $G$ FLSs such that (1) an FLS participates in exclusively one group and all FLSs agree on participation in the same group, and (2) the euclidean distance between the group members is minimized.
\end{problem}

The symmetry constraint that all FLSs agree to be a member of the same group is fundamental.
It defines a maximum for the number of groups:
$\lfloor \frac{F}{G} \rfloor$.
A heuristic may produce {\em ceded} FLSs.
These FLSs belong to a group with fewer than $G$ FLSs; potentially a group with the FLS by itself.
The theoretical lower bound on the number of ceded FLSs is $F \% G$.
While the grouping problem has a polynomial time solution with $G$=2~\cite{g=2}, it is NP-hard with $G \geq 3$~\cite{groupingNPhard}.

\noindent{\bf Optimality and Outring:}
To evaluate the alternative decentralized group formation techniques, we developed Outring, a scalable technique that constructs swarms of FLSs with well known solutions.
With some settings of Outring, the known solution is optimal.
With others, it is one solution out of many possibilities.
Outring works in both 2D and 3D.
Below, we describe it in 2D because it is simpler to visualize.

\begin{figure}
\centering
\includegraphics[width=\columnwidth]{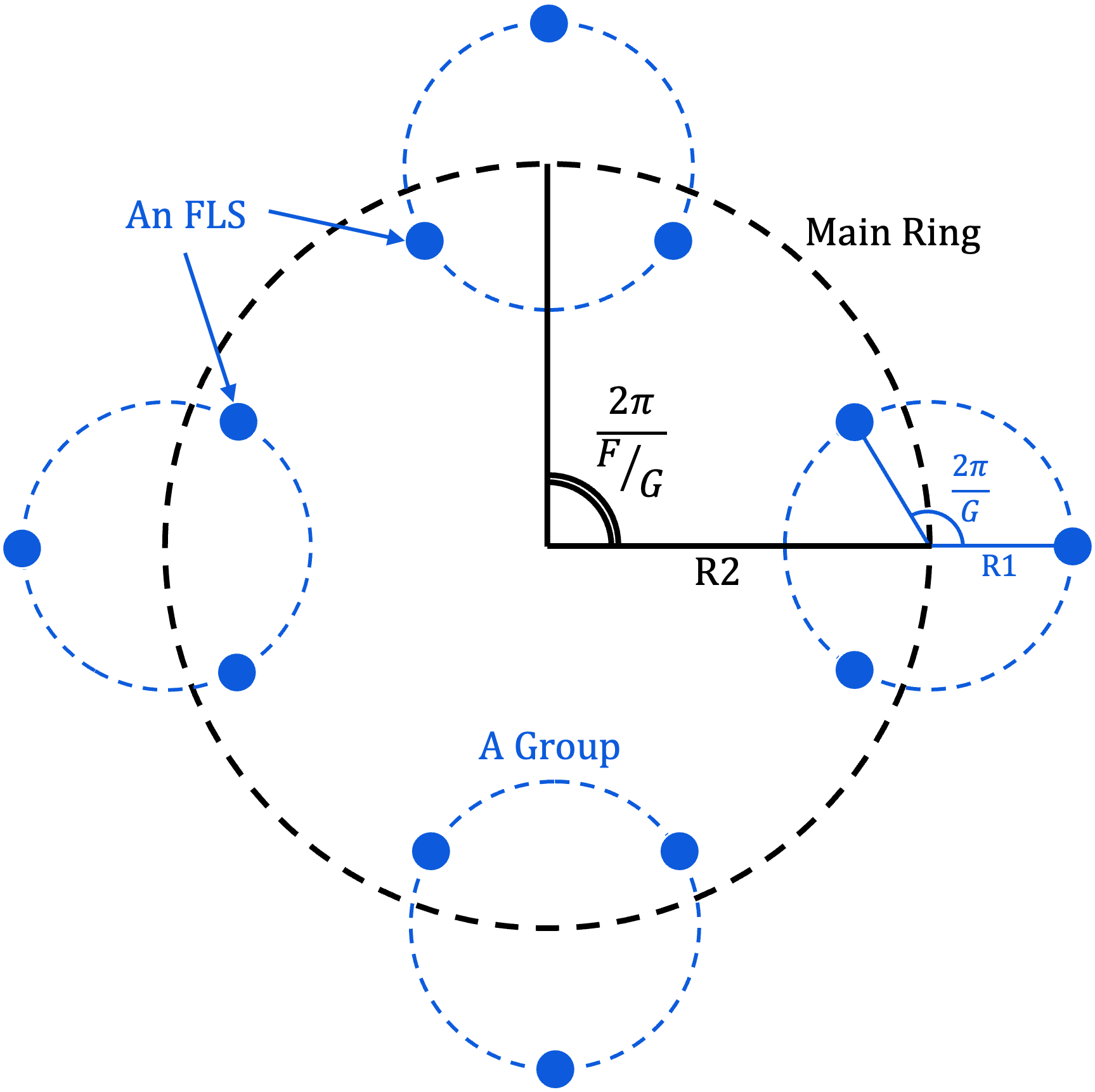}\hfill
\caption{An Outring with F=12 FLSs and G=3 FLSs/group.}
\label{fig:outring}
\end{figure}

Outring assumes $F$ is a multiple of $G$ ($F \% G$=0) and constructs $\frac{F}{G}$ groups on the circumference of a main ring with radius R2, see Figure~\ref{fig:outring}.
Each group is a ring with radius R1.
The G FLSs that constitute a group are placed equally apart on the circumference of its ring.
The center of these circles (groups) are equally spaced apart on the circumference of the main ring.
Figure~\ref{fig:outring} shows the Outring with 12 FLSs and 4 groups.
The identified angle for a group dictates the location of its G FLSs.
The ratio $\rho=\frac{R2}{R1}$ dictates how far apart the groups are on the circumfrance of the main ring.
A high value of $\rho$, say 100, results in a sparse point cloud with groupings that are trivial to identify by a human.
With this setting of Outring, the optimal minimum ($d_{min}$) and maximum distance ($d_{max}$) between a pair of FLSs in a group is:
\begin{equation}\label{eq:min}
    d_{min}(G)=R_1\sqrt{2(1-cos\frac{2\pi}{G})}
\end{equation}

\begin{equation}
    d_{max}(G)=\begin{cases}
        2R_1, & \text{if $G$ is even}.\\
        R_1\sqrt{2(1+cos\frac{\pi}{G})}, & \text{otherwise}.
    \end{cases}
\end{equation}
The average distance, $d_{avg}$, between the FLSs that constitute a group is:
\begin{equation}\label{eq:avg}
    d_{avg}(G)=\begin{cases}
         \frac{4R_1}{G-1}( \frac{1}{2}+ \sum_{i=1}^{\frac{G}{2}-1} sin\frac{i\pi}{G}) , & \text{if $G$ is even}.\\
         \frac{4R_1}{G-1} \sum_{i=1}^{\frac{G-1}{2}} sin\frac{i\pi}{G}, & \text{otherwise}.
    \end{cases}
\end{equation}
A small setting for $\rho$ (say 1) produces a dense point cloud, causing the FLSs that constitute different groups to overlap one another.
In these scenarios, Equations~\ref{eq:min}-~\ref{eq:avg} are not optimal.
They simply describe a solution out of many.  




\section{Decentralized Group Formation}\label{sec:decentralized}
This section describes several decentralized heuristics to form groups.
An FLS $f_i$ executes a heuristic by itself upon arriving at the coordinate assigned to it for illuminating a point~\cite{shahram2022}.  
A heuristic requires the neighbors of $f_i$, $N(f_i)$.
This section presents the heuristics assuming $N(f_i)$ is provided to each $f_i$.
Section~\ref{sec:impl} describes how an FLS $f_i$ discovers its neighbors.  
Below, we adapt the decentralized techniques of~\cite{Chmielowiecvns2010,kclique2014} to construct groups.

We define $\Delta(f_i,f_j)$ as the distance between two FLSs $f_i$ and $f_j$,
$\Delta(f_i,f_j)=\Delta(f_j,f_i)$.
It may be computed using the 3D coordinates of the point assigned to each of $f_i$ and $f_j$.
Each FLS $f_i$ maintains a group $R_i$ consisting of $G-1$ unique FLSs. 
The weight of any two FLSs $f_p$ and $f_q$ in $R_i$ is the reciprocal of their distance:
$w(f_p,f_q)=\frac{1}{\Delta(f_p,f_q)}$.
This definition of weight is symmetric, $w(f_p,f_q)=w(f_q,f_p)$.
The objective of the decentralized technique is for an FLS $f_i$ to compute set $R_i$ that maximizes the sum of the weights between any 2 FLS in its group\footnote{This equation assumes the FLS $f_i$ is added to its group $R_i$, increasing its cardinality to $G$.}:

\begin{equation}\label{eq:weight}
W_i(R)=\sum_{(f_p, f_q)\in R \cup \{f_i\}}^{}w(f_p,f_q).
\end{equation}

We require the decentralized technique to terminate.
This means all FLSs must stop forming new groups after sometime.  It is possible that the decentralized technique terminates with some FLSs not forming a group of $G$ FLSs.
For each such {\em ceded} FLS $f_p$, its group is either the empty set or consists of a set of FLSs that do not agree on the same grouping.

A group $R=\{f_1, \cdots, f_{G-1}\}$ is {\em proper}~\cite{kclique2014} if and only if for each neighbor $f_j$ this set has a higher weight than all other groups found by that neighbor.
If this is true for all FLSs of $R$ then
the predicate proper($f_i$, $R$) returns true.
A score for a set $R$ and FLS $f_i$ is defined as its weight $W_i(R)$ if the set is proper.  Otherwise, its score is zero.
\begin{equation}\label{eq:score}
    Score_i(R)=\begin{cases}
        W_i(R), & \text{If proper($f_i$, $R_i$) is true}.\\
        0 & \text{Otherwise}.
    \end{cases}
\end{equation}

To compute $Score_i(R)$ of Equation~\ref{eq:score}, an FLS $f_i$ exchanges messages with its neighbors, N($f_i$).
Its message contains its $W_i$.
FLS $f_i$ gathers this information from the messages transmitted\footnote{These transmissions facilitate discovery of neighboring nodes, see Section~\ref{sec:impl}.} by its neighbors to compute its $Score_i(R)$.

We define the convergence of a decentralized techniques using either one of the following two conditions:
1) All FLSs are reporting the same grouping, 
2) The grouping computed by all FLSs does not change for a fixed amount of time $\Delta$.
$\Delta$ is set to 2 minutes in our experiments.
If either condition holds true then the FLSs decide the technique has converged and stop executing the technique.

The formal definition of the first termination condition is as follows:
\{$\forall p \in R_i: R_p \cup \{f_p\} = R_i \cup \{f_i\} ~and~ |R_p|=|R_i|=G-1 \}$.

A basic decentralized algorithm may require an FLS $f_i$ to evaluate  all ${|N(f_i)| \choose G-1}$ possible combination of forming a group from its neighbors $N(f_i)$~\cite{Chmielowiecvns2010,kclique2014}.
It maintains the one with the highest score and sends the new group and its weight to all neighbors.
A receiving neighbor  updates its group to include $f_i$ only if it is included in $f_i$'s $R_i$.
This algorithms is shown to be slow when an FLS has many neighbors~\cite{Chmielowiecvns2010,kclique2014}.
This is a common occurrence with dense point clouds, e.g., wheels of the Skateboard in Figure~\ref{fig:4_shapes}.c. 
Heuristics speed up this basic algorithm.

\subsection{Heuristics}
Below, we describe Random and several variants of it including RS~\cite{Chmielowiecvns2010,kclique2014}, Variable Neighborhood Search (VNS)~\cite{vns2009,Chmielowiecvns2010}, and Closest Available Neighbor First (CANF).
A technique may have a configuration parameter.
For example, VNS requires the application to specify the amount of time it spends in its local search.

\subsubsection{Simple Random, SimpleR}\label{sec:simpler}  
In its simplest, Random requires an FLS $f_i$ to select G-1 FLSs from a fixed number of neighbors, $N(f_i)$.
We term this technique SimpleR.
The FLSs in $N(f_i)$ are sorted in ascending order of their distance from $f_i$ with the closest FLS in the Euclidean space appearing first. 
For a given group size $G$, we consider the following cardinalities of $|N(f_i)|$:  $G-1$, $G$, and $1.5 \times G$.
Setting $|N(f_i)|$=$G-1$ requires an FLS to consider only its closest neighbor in the Eucledian space.
An evaluation of the settings using the Outring shows $|N(f_i)|$ set to $1.5 \times G$ is typically inferior to the other two settings.

\subsubsection{Random Subset, RS}\label{sec:rs} 
RS~\cite{kclique2014} is a variant of random.  It selects a random set of FLSs from its known neighboring FLSs ($N(f_i)$) and considers all possible $G-1$ groupings to select the highest scoring one.
SZ is an input parameter that dictates the number of FLSs that RS must permutate,
SZ $\geq G-1$.
If the number of neighbors known to an FLS
is fewer than SZ then RS returns the empty set, forcing the framework to wait until there are enough neighbors.  
An FLS that receives a message broadcasted by its neighboring FLSs will add them to its neighbor list.
This is useful with sparse point clouds, enabling RS to form small groups quickly.

Once the neighbor list of an FLS consists of at least SZ FLSs\footnote{With the radio range of each FLS set to the entire display, $f_i$ populates its $N(f_i)$ with all FLSs quickly.},
RS selects SZ elements of it randomly,
enumerating all possible combinations of this set with those in the FLS's current group.
It iterates this enumeration to identify the highest scoring one\footnote{In our implementation, RS computes Score by caching the grouping information provided by each neighboring FLS.  There is no network communication in computing scores.}.

In our Outring experiments, RS computes the optimal grouping with small group sizes, $G\leq$5.
With larger groups ($G$=15) and a dense Outring ($\rho$=1),
RS fails to form even one group. 
The probability of an FLS choosing a subset of its neighbors to form a proper group is low.
This probability is 
$\frac{ {F-1-(G-1) \choose SZ-(G-1)} }{ {F-1 \choose SZ} }$.
With F=90 FLSs and $G$=15, this probability is the astronomically low value of 4.19e-11.
In order for this probability to increase above 0.001 with $G$=15, one must reduce the number of FLSs to fewer than 33.

\subsubsection{Variable Neighborhood Search, VNS}\label{sec:vnc}

VNS~\cite{vns2009,Chmielowiecvns2010,kclique2014} solves the heaviest subgraph problem where each subgraph consists of $G$ vertices (FLSs).  The edge between two vertices $f_p$ and $f_q$ denotes the weight between them, see definition of $w(f_p,f_q)$.  We consider this heuristic because it has been applied in diverse scenarios including data mining, operations research, placement of warehouses, knapsack and packing, and scheduling.  It takes a global approach to forming groups by identifying a subset of the solution space that is close to a particular point in the global solution space.  Subsequently, it refines this solution by performing a local search.   

VNS defines the {\em d-th neighborhood structure} of a certain grouping $C$ to consist of all grouping that differ by at most $d$ FLSs from $C$.  It consists of two functions, Shake and LocalSearch, that are executed in turn.  The function Shake avoids local optimum.  It modifies the current grouping C by randomly changing a few of its FLSs to generate a C’ that belongs to the d-th neighborhood structure.  The LocalSearch function improves upon C’ by considering all possible FLSs known to $f_i$, keeping the highest scoring one.

A challenge of VNS is the amount of time allotted for the Shake and LocalSearch operations.
A small amount of time, say 10 milliseconds, enables VNS to compute small groups ($G$=3) with sparse point clouds ($\rho$=100) quickly.  
However, the same settings with a dense point cloud ($\rho$=1) and large group sizes ($G$=10) requires more than an hour of execution time. 
In this case, allotting a longer amount of time results in a significantly faster execution time.
We evaluated 10, 40, 100, and 1000 milliseconds as the amount of alloted time.  
In general, 100 milliseconds provides a competitive response time with high quality groupings.

\subsubsection{Closest Available Neighbor First, CANF}\label{sec:canf}

CANF explores the neighborhood of an FLS $f_i$ to construct a group.
It sorts its neighboring FLSs in ascending distance using their assigned coordinates.
It iterates this list starting with the closest FLS $f_u$.
If $f_u$ does not have $f_i$ as a group member, CANF evaluates each of $f_u$'s group members as its own.
It uses its local $R_u$ information about $f_u$ to construct a group and decide if its grouping is proper and compute a score.  
It adds $f_u$ as a candidate only if it results in a higher score.

CANF is able to find the optimal solutions in all scenarios.
With small values of $G$, its response time is faster than 40 milliseconds.
A surprising result is its sweet spot with $G$=5, providing response times faster than $G$=3.  

\section{An Implementation}\label{sec:impl}
\begin{table*}[h]
\caption{Two sparse Outring point clouds and group sizes, $F$=90, $\rho$=100, 1 Amazon AWS instance with 96 physical cores.}\label{tbl:cmpsparse}
\begin{small}
\begin{tabular}{|c||c|c|c|c||c|c|c|c|}
\hline
\hline
 Sparse & \multicolumn{4}{c||}{$G$=3} & \multicolumn{4}{c|}{$G$=15}  \\ 
 \cline{2-9}
\cline{2-9}
$\rho$=100 & CANF & SimpleR:$|N(f_i)|$=G & VNS (100 msec) & RS:$SZ$=G-1 & CANF & SimpleR:$|N(f_i)|$=G & VNS (100 msec) & RS:$SZ$=G-1 \\
\hline
\hline
RT (Sec) & 0.041 & 0.23 & 0.64 & 2.19 & 0.095 & 0.30 & 233.3 & 262.8 \\
\hline 
$nG$ & 30 & 30 & 30 & 30 & 6 & 6 & 3.875 & 3.867 \\
\hline
\# Ceded FlSs & 0 & 0 & 0 & 0 & 0 & 0 & 31.875 & 32 \\
\hline
Avg Distance & 1.73 & 1.73 & 1.73 & 1.73 & 1.36 & 1.36 & 1.36 & 1.36 \\ 
\hline
Xmt BW (Gbits/Sec) & 1.57 & 0.17 & 0.0083 & 0.075 & 0.3 & 0.03 & 0.000040 & 0.0016 \\ 
\hline
Rec BW (Gbits/Sec) & 0.43 & 7.58 & 0.22 & 3.87 & 1.21 & 0.97 & 0.0019 & 0.0038 \\ 
\hline
\# of Executions & 121 & 102 & 16 & 543 & 72 & 23 & 30 & 34332 \\
\hline
\hline
\end{tabular}
\end{small}
\end{table*}

\begin{table*}[h]
\caption{Two dense Outring point clouds and group sizes, $F$=90, $\rho$=1, 1 Amazon AWS instance with 96 physical cores.}
\label{tbl:cmpdense}
\begin{small}
\begin{tabular}{|c||c|c|c|c||c|c|c|c|}
\hline
\hline
 Dense & \multicolumn{4}{c||}{$G$=3} & \multicolumn{4}{c|}{$G$=15}  \\ 
 \cline{2-9}
\cline{2-9}
$\rho$=1 & CANF & SimpleR:G & VNS (100 msec) & RS:$SZ$=G-1 & CANF & SimpleR:G & VNS (100 msec) & RS:$SZ$=G-1 \\
\hline
\hline
RT (Sec) & 121.05 & 120.60 & 189.88 & 123 & 372.65 & 320.02 & 4305.54 & 233.54 \\ 
\hline
$nG$ & 26 & 22 & 25 & 25 & 3.625 & 1.13 & 3 & 0 \\ 
\hline
\# Ceded FlSs & 12 & 24 & 15 & 15 & 35.625 & 73.13 & 45 & 90 \\ 
\hline
Avg Distance & 0.32 & 0.29 & 0.32 & 0.32 & 0.98 & 0.77 & 0.86 & 1.45 \\ 
\hline
Xmt BW (Gbits/Sec) & 0.09 & 0.12 & 0.003 & 0.064 & 0.003 & 0.03 & 0.000015 & 0.000026 \\ 
\hline
Rec BW (Gbits/Sec) & 4.04 & 3.93 & 0.05 & 3.7 & 0.07 & 1.54 & 0.00127 & 0.00056 \\
\hline
\# of Executions & 27,693 & 36,882 & 1,710 & 25,893 & 2,896 & 30,363 & 211 & 9 \\
\hline 
\hline
\end{tabular}
\end{small}
\end{table*}

An FLS process consists of two threads, a networking ($T_{Network}$) and a handler ($T_{Handler}$) thread.  These two threads communicate using a shared queue.  $T_{Network}$ listens on a UDP socket, receives a message, generates an event in response to the messages, and inserts the events in the shared queue for processing by $T_{Handler}$.  $T_{Handler}$ blocks on the queue for an event.  After receiving the event, it invokes a method to process the event.  Subsequently, it inserts an event in the queue that causes it to execute again.  This implements an infinite while loop.
A specific event terminates $T_{Handler}$.

Except for CANF, other heuristics require an FLS process to use its maximum radio range to communicate with all other FLSs.  CANF starts the radio range of an FLS to be one display cell from its current location.  When an FLS does not have enough neighbors to form a group consisting of G-1 FLSs, it expands its radio range by a factor of two.  This FLS does not expand its radio range again for a pre-configured interval of time.  In our experiments, this interval is set to 50 milliseconds. 

Periodically, an FLS broadcasts the following metadata to all FLSs in its radio range:  its unique identifier $f_i$, coordinates,  group and its score,
and a monotonically increasing message id.
The latter enables a receiver to detect out of order messages and to discard them.  Messages may arrive out of order due to our use of UDP.

A receiving FLS builds a profile of its neighbors using a hash table.
Its key is its neighbor's FLS id, $f_i$, and its value is the received metadata.
To implement a technique such as SimpleR, an FLS sorts its neighbors using their coordinates and truncates the list based on its setting $G-1$, $G$, or $1.5 \times G$.


Our implementation consists of a primary process and many secondary processes.  These processes execute on a cluster of multi-core servers.
The primary process executes on one server and launches a secondary process on each server.
The servers are identified to the primary process using a configuration file.
Primary and secondary processes start one or more FLS processes on their servers. 


Each secondary process opens an administrative network socket to the primary process.
The primary polls each secondary every 40 milliseconds to detect either of the two convergence conditions stated in Section~\ref{sec:decentralized}.
Once a termination condition is satisfied, it generates a message that causes $T_{Network}$ and $T_{Handler}$ to terminate (stop generating events in its queue), stops the experiment, and provides a log file that contains the results.

\section{A Comparison}\label{sec:cmp}

This section compares the alternative heuristics using an Outring and the Chess point cloud of Figure~\ref{fig:4_shapes}.a.
The Outring studies were conducted using an Amazon AWS instance c6a.metal with 384 Gigabytes of memory and 96 physical cores (192 virtual core).  The clock speed of each core is 3.6 GHz.
The Chess point cloud experiments were conducted using a cluster of 16 CloudLab servers, each with 256 GB of memory, two 32-core 2.8 GHz hyperthreaded CPUs (model r6525, Clemson data center).
With both, each FLS process runs on a physical core.

\noindent{\bf Outring.}
Tables~\ref{tbl:cmpsparse} and~\ref{tbl:cmpdense} provide a comparison of different heuristics with a sparse ($\rho$=100) and a dense ($\rho$=1) Outring consisting of 90 FLSs ($F$=90).  
In general, CANF maximizes the number of constructed groups ($nG$) and computes these groups faster (response times, RT) than the other techniques.
However, it uses more system resources in the form of network bandwidth, see the average bandwidth required by each FLS to transmit data.   

With a sparse Outring ($\rho$=100) and small group size ($G$=3), all heuristics find the optimal solution, computing 30 groups ($\frac{F=90}{G=3}$) with no ceded FLSs.  
The same is not true with a dense Outring ($\rho$=1).  While the optimal solution is not known for this synthetically generated point cloud, the heuristic that maximizes the number of groups ($nG$) with competitive response time (RT) and average distance between FLSs that constitute a group is superior. 
CANF qualifies these requirements. 
A limitation of CANF is its higher bandwidth requirement than the other techniques.

All techniques become slower with a larger group size, $G=15$.
CANF and SimpleR provide the fastest response times with the sparse Outring ($\rho$=100), see Table~\ref{tbl:cmpsparse}.  They compute almost twice as many groups when compared with VNS and RS.  The distance between FLSs in a group is the optimal value.  CANF is twice as fast as SimpleR.  However, its average transmission bandwidth for each FLS is also twice as high.
A dense Outring ($\rho$=1) slows down all heuristics considerably, see Table~\ref{tbl:cmpdense}.
CANF continues to maximize the number of found groups ($nG$).
However, it fails to construct six groups.
At the same time, it is an order of magnitude faster than VNS.
SimpleR and RS are faster than CANF. 
However, they fail to construct a single group in many experiments.  


\begin{table*}[h]
\caption{A comparison of the alternative decentralized techniques using the Chess piece point cloud requiring 454 FLSs.}\label{tbl:chess}
\begin{small}
\begin{tabular}{|c|c|c|c|c||c|c|c|c|}
\hline
\hline
 & \multicolumn{4}{c||}{G=3} & \multicolumn{4}{c|}{G=20}  \\ 
 \cline{2-9}
 & CANF & SimpleR:$G$ & VNS (100 msec) & RS:SZ=G-1 & CANF & SimpleR:$G$ & VNS (100 msec) & RS:SZ=G-1 \\ 
\hline
\hline
    RT (Sec) & 51.23 & 126.7 & 204.2 & 36.43 & 4,574 & 200.25 & 99,153 & 166.33 \\ 
    \hline
    nG & 151 & 122 & 151 & 151 & 21 & 17.13 & 9 & 0\\
    \hline
    \# Ceded FLSs  & 1 & 88 & 1 & 1 & 34 & 111.5 & 274 & 454 \\
    \hline
    Avg Distance & 5.77 & 4.8 & 5.59 & 5.47 & 13.13 & 9.9 & 9.24 & 36.24 \\ 
    \hline
    Xmt BW (Gbits/Sec) & 0.39 & 0.45 & 0.014 & 0.46 & 0.003 & 0.07 & 0.000005 & 0.0000016 \\ 
    \hline
    Rec BW (Gbits/Sec) & 21.2 & 10.13 & 0.49 & 14.13 & 0.13 & 2.56 & 0.0026 & 0.0002206 \\ 
    \hline
    \# of Executions & 9,529 & 37,392 & 1,419 & 8,419 & 6,801 & 9,477 & 279 & 11 \\
    
\hline
\hline
\end{tabular}
\end{small}
\end{table*}

\noindent{\bf Chess Point Cloud.}
Table~\ref{tbl:chess} compares the performance of the alternative techniques with a small ($G$=3) and a large ($G$=20) group size using the Chess piece of Figure~\ref{fig:4_shapes}.a.

With $G$=3, RS is the superior technique.  
It provides the fastest response time while maximizing the number of groups.
This trend does not hold true across all shapes.
With the Skateboard, CANF is the fastest technique (4x faster than RS) and provides the highest number of groups.
In most cases, CANF requires a higher network bandwidth.
It is as high as 41 Gbits/sec with the Race car.

With $G$=20, CANF maximizes the number of groups.
However, it is 20x slower than SimpleR:G that computes some groups.
Even though RS is the fastest technique, it fails to construct a single group.
SimpleR:G requires the highest amount of bandwidth.
These trends hold across different shapes in Figure~\ref{fig:4_shapes}.

\begin{figure}[h]
\centering
\includegraphics[width=\columnwidth]{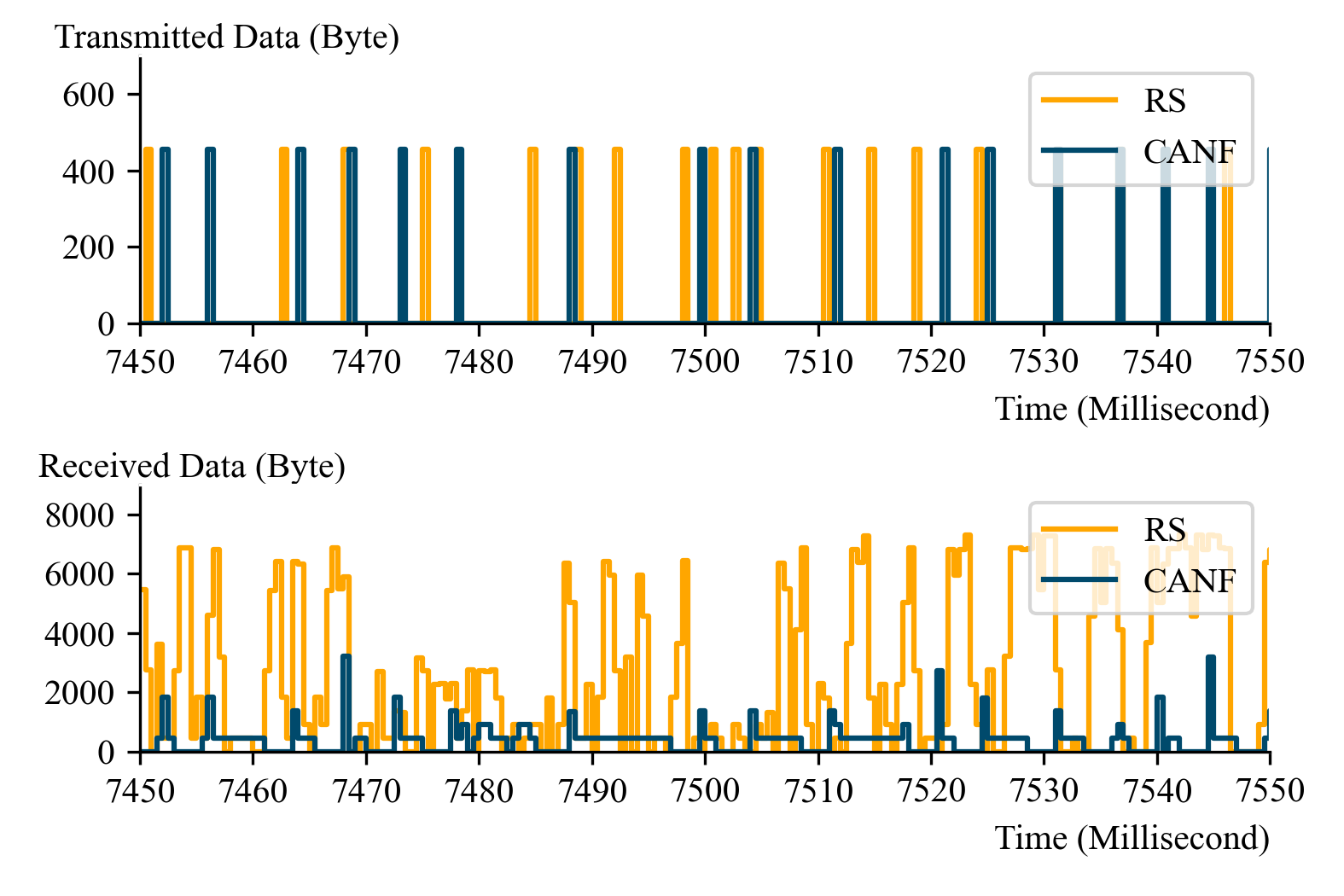}\hfill
\caption{Bytes transmitted and received by one FLS, G=3.}
\label{fig:bw_comp_canf_rs}
\end{figure}

Figure~\ref{fig:bw_comp_canf_rs} shows the number of bytes transmitted and received by each FLS using RS and CANF as a function of time.  The required bandwidth is bursty as a function of time.  Most of the network traffic is driven by the 20\% probability that requires an FLS to execute a heuristic even though there is no change in the neighbors or the discovered groups.  The rate at which data is received is different because CANF expands its radio range incrementally while RS requires an FLS to use its maximum radio range.

The results of Table~\ref{tbl:chess} motivate a hybrid heuristic 
using both RS and CANF.
See Section~\ref{sec:conc}.



\section{Related Work}\label{sec:related}
The concept of FLS displays and illuminations is presented in~\cite{shahram2021,shahram2022,mmsys2023}.
Use of a reliability group with one or more standby FLSs to tolerate FLS failures is described in~\cite{shahram2022}.
It quantifies the benefits of a reliability group assuming they are already constructed.
It does not describe how FLSs may form groups. 
It may use the decentralized algorithms described in this study to form groups.

Forming groups is a multi-disciplinary research topic.
Its roots in graph theory date back to 1947~\cite{tutte47}.
The challenge is termed matching~\cite{edmonds65} or weighted matching~\cite{avis83}, H-packing~\cite{hpacking2005,hpack2007,hpack2008}, H-matching~\cite{kann1994,kann1997}, H-partition~\cite{groupingNPhard,hpart2020}, k-subgraph~\cite{vns2009}, k-clique~\cite{Chmielowiecvns2010,kclique2014}, clustering~\cite{kmeans2007,lloyd1982,kmean2022} among others. 
There are subtle and yet fundamental differences between the alternative studies.
For example, in the context of peer-to-peer networks, a protocol may allow a node to serve as a preferred neighbor for any number of other nodes~\cite{tman2009,voulgaris2006} without requiring those nodes to reciprocate the neighbor relationship.
This is different than our problem that requires an FLS to be a member of one group and all FLSs in the group to agree to be a part of the group.

A comparison of VNS~\cite{vns2009}, RS~\cite{Chmielowiecvns2010}, and a basic exhaustive search using a simultation study is presented in~\cite{Chmielowiecvns2010,kclique2014}.
This study uses a simple complete graph where each node has every other node as a neighbor with a random value between (0,1) as the weight of the edge.
One conclusion of this study is that RS and VNS are superior to the basic exhaustive search by maximizing the number of groups.  It also concludes that RS is slightly faster than VNS and favors RS for its simplicity.
Our evaluation is different in that we use real point clouds with a real implementation.
We compare the different heuristics with group sizes as large as 20 (G$\leq$20).
Our results are consistent with those of~\cite{Chmielowiecvns2010,kclique2014} for small group sizes (G$\leq$5).
However, we find RS may compute very few groups with large group sizes, G$\geq$10.
VNS computes more groups and is significantly slower than RS.
Another novelty of our study is 
the introduction of 
SimpleR and CANF.
Obtained results show both are superior to the other techniques with large group sizes, G$\geq$10.  


\section{Conclusions and Future Research}\label{sec:conc}
This study evaluates four decentralized techniques to form groups of size $G$.
These include SimpleR, RS~\cite{Chmielowiecvns2010}, VNS~\cite{vns2009,Chmielowiecvns2010}, and CANF.
Obtained results show
RS is superior to the others with small values of $G$ ($\leq$ 5).
With large group sizes ($\geq$ 10), either CANF or SimpleR:G are superior depending on whether the objective is to maximize the number of groups or compute some groups quickly.
This motivates a hybrid technique that uses the appropriate technique given a desired number of groups, $nG$.
It is a short term research direction.

The topology of a point cloud dictates the response time of a technique with different values of $G$.
Generally, a technique becomes slower with larger point clouds and group sizes, i.e., higher values of $G$.  
However, this is not true always.
A point cloud such as the Skateboard of Figure~\ref{fig:4_shapes}.c may consist of more than 2x (4x) points than the Dragon (Chess piece).  Yet, a technique such as CANF executes 10x (7x) faster with the Skateboard than the Dragon (Chess piece).  Similarly, we observed CANF with $G$=5 to be faster than with $G$=3 using both the Outring and the Race car of Figure~\ref{fig:4_shapes}.d.
Finally, we observed network packet loss as high as 10\% has minimal impact on the performance of CANF.


In many cases, a heuristic forms different sized groups.
An interesting research direction is to form groups of larger size incrementally.
An example is a haptic interaction that exerts an increasing amount of force on the user as they press further into a virtual object~\cite{salisbury2004haptic}.
The display may use the smaller FLS groups to exert a small amount of force while larger groups join in waves to increase the amount of exerted force. 

\begin{acks}
This research was supported in part by the NSF grant IIS-2232382.  We gratefully acknowledge CloudBank~\cite{cloudbank2021} and CloudLab~\cite{emulab} for the use of their resources to enable all experimental results presented in this paper.
\end{acks}


\bibliographystyle{ACM-Reference-Format}
\bibliography{refs}


\end{document}